\documentclass[twocolumn,amsmath,amssymb]{revtex4}
\usepackage[dvipdfmx]{graphicx}
\usepackage{here}
\usepackage{epsfig}
\usepackage{times}
\usepackage{amsmath}
\usepackage{comment}
\usepackage{amsfonts}
\usepackage{amssymb}
\usepackage[usenames]{color}
\usepackage{fouriernc}
\usepackage[dvipdfmx]{hyperref}

\newcommand{\NTT}{NTT Basic Research Laboratories, NTT Corporation, 3-1 Morinosato-Wakamiya, Atsugi, Kanagawa 243-0198, Japan.}
\newcommand{\KYOTO}{Institute for Chemical Research, Kyoto University,
Gokasho, Uji, Kyoto 611-0011, Japan}
\newcommand{\keio}{School of Fundamental Science and Technology,
Keio University, 3-14-1 Hiyoshi, Kohoku-ku, Yokohama 223-8522, Japan}
\newcommand{\AIST}{Correlated Electronics Group, Electronics and Photonics Research Institute, National Institute of Advanced Industrial Science and Technology (AIST), Tsukuba Central 5, 1-1-1 Higashi, Tsukuba, Ibaraki 305-8565, Japan}
\newcommand{\CERN}{Center for Spintronics Research Network, Keio University, 3-14-1 Hiyoshi, Kohoku-ku, Yokohama 223-8522, Japan}

\newcommand{\ket}[1]{\ensuremath{\vert#1\rangle}}
\newcommand{\bra}[1]{\ensuremath{\langle #1\vert}}

\begin{document}

\title{Bandwidth analysis of AC magnetic field sensing based on electronic spin
double resonance of nitrogen-vacancy centers in diamond}

\author{Tatsuma Yamaguchi}
   \affiliation{
\keio
   }
\author{Yuichiro Matsuzaki}
   \affiliation{
\NTT
   }
\author{Shiro Saito}
   \affiliation{
 \NTT
   }
 \author{Soya Saijo}
   \affiliation{
\keio
   }
\author{Hideyuki Watanabe}
   \affiliation{
\AIST
   }
\author{Norikazu Mizuochi}
   \affiliation{
\KYOTO
   }
\author{Junko Ishi-Hayase }
  \affiliation{
\keio \\ \CERN
   }

\begin{abstract}Recently we have demonstrated AC magnetic field sensing scheme using a simple continuous-wave optically detected magnetic resonance of nitrogen-vacancy centers in diamond [Appl. Phys. Lett. \textbf{113}, 082405 (2018)]. This scheme is based on electronic spin double resonance excited by continuous microwaves and radio-frequency (RF) fields. Here we measured and analyzed the double resonance spectra and magnetic field sensitivity for various frequencies of microwaves and RF fields. As a result, we observed a clear anticrossing of RF-dressed electronic spin states in the spectra and estimated the bandwidth to be approximately 5 MHz at the center frequency of 9.9 MHz.
\end{abstract}

\maketitle
Nitrogen-Vacancy (NV) center is a defect in diamond where two adjacent carbon atoms are replaced by a nitrogen and a vacancy. It is possible to initialize the electronic spin states of negatively-charged NV centers by illuminating them with green laser\cite{Harrison2004}. 
Also, photoluminescence from the NV centers provides us with a way to readout the spin states\cite{Gruber1997,Jelezko2002}. Moreover, the spin states can be manipulated with microwave pulses\cite{Jelezko2004}, and the coherence time of the NV center is as long as 2 ms even at a room temperature\cite{Balasubramanian2009}.
These properties show a potential of the NV centers as a high performance magnetic field sensor\cite{Taylor2008,Wolf2015a,Balasubramanian2008}.There are many applications such as vector magnetic field 
sensing\cite{Maertz2010,Steinert2010,Pham2011,Tetienne2012,Dmitriev2016,Kitazawa2017,Schloss2018,Zhang2018,Wang2015,Yahata2019} and magnetic field imaging\cite{hatano2018,LeSage2013,Pham2011,Steinert2013,Glenn2015,Devience2015} with NV centers.\par

In the conventional AC magnetic field sensing, pulsed-optically detected magnetic resonance (pulsed-ODMR) has been used\cite{Taylor2008,Wolf2015a}. Although pulsed-ODMR enabled us to achive high sensitivity, it requires sophisticated calibrations for the pulse control. In particular,  
as the frequency of the AC magnetic fields increases, it becomes technically more difficult to detect the AC magnetic fields in the conventional pulsed ODMR scheme because of the requirement of the shorter pulse intervals (〜ns). 
Futhermore, when AC magnetic field sensing  is performed with the pulsed-ODMR that requires high-speed measurements,  it is not straightfoward to use a wide field imaging with a CCD camera that has a slow response\cite{Pham2011}.\par

Recently, our group has proposed and demonstrated novel technique of AC magnetic field sensing with the continuous wave-ODMR (CW-ODMR), which dose not require the pulse control and high-speed measurements. This scheme is based on
double resonance of sub-levels of spin-triplet state of NV centers excited by simultaneous application of microwaves and radio-frequency (RF) fields\cite{saijo2018}.
The microwaves play a role of probing NV centers dressed by the RF fields which correspond to the target AC magnetic fields. In the ground state manifolds of the NV centers, there are three states such as  
\ket{B}, \ket{D}, \ket{0}. These states are eigenstates of spin-triplets of NV centers under the application of stress/electric field. When the RF field is resonant with the transition between \ket{B} and \ket{D}, there are changes in the ODMR spectrum with a sweep of the microwave frequency.  
Such changes in the ODMR signals allows us to detect the AC magnetic fields.\par

However, the AC magnetic field sensing with CW-ODMR was demonstrated with a fixed frequency that is resonant with the transition between \ket{B} and \ket{D} in \cite{saijo2018}. For a practical purpose, 
it is important to determine the bandwith of the AC magnetic field sensor. In this paper, we have investigated the frequency dependence of the sensitivity with the AC magnetic field sensor using CW-ODMR. 
Firstly, we have performed a double resonance experiment with the CW-ODMR where we sweep the frequencies of the microwave and radio frequency. In the spectrum, we clearly observed anticrossing which corresponds to Autler-Townes (AT) splitting induced by RF field. Secondly, we have well reproduced the experimental results with theoretical calculations.
Finally, we have estimated the sensitivity as the AC magnetic field sensor with several frequencies by using both experimental and theoretical results, and we have confirmed that there is a good agreement between them.
\par

\begin{figure}[htbp]
\begin{center}
\includegraphics[width=7.5cm]{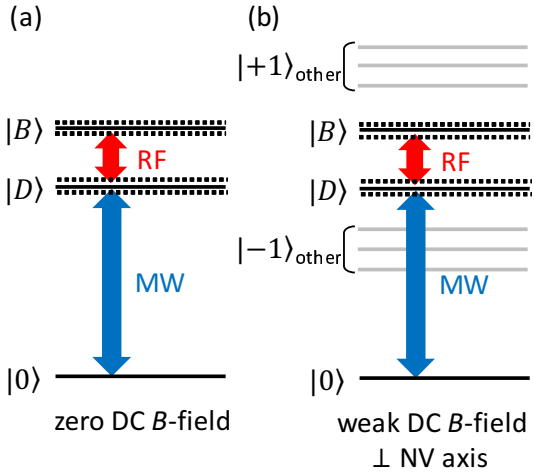}
\caption{Energy levels of the NV centers. (a) Without external DC magnetic fields, NV centers with different crystallographic axes have the same resonant frequency, and there are two MW resonances in the spectrum. Under the resonant excitation by RF field, these are four MW resonance (dotted line) which corresponds to AT splitting caused by RF field.
(b) With applied DC magnetic fields perpendicular to one of the four possible crystallographic axes, the degeneracy between the axes is lifted, and there are eight MW resonances in the spectrum.
We focus on the NV centers with an axis perpendicular to the applied DC magnetic fields by the frequency selectivity.
}
\abovecaptionskip=12pt
\label{f:NVstate}
\end{center}
\end{figure}

In our experiment, we apply external magnetic fields to obtain a clear spectrum while the previous demonstration of the AC magnetic field sensing with CW-ODMR was done without applied magnetic fields \cite{saijo2018}. 
NV centers have four possible crystallographic axes, and 
we apply magnetic fields perpendicular to one of the crystallographic axes. The magnetic fields lift the degeneracy of the NV centers so that the resonant frequency of a quarter of the NV centers can be different from the other NV centers as shown in Fig. \ref{f:NVstate}. 
Importantly, with applied magnetic fields of a few mT perpendicular to the crystallographic axis, the energy eigenstates are still approximately described by \ket{0}, \ket{B}, and \ket{D}. 
So the AC magnetic fields can induce the transition between \ket{B} and \ket{D}, and this let us perform the AC magnetic field sensing with CW-ODMR.

Let us explain the theoretical part of our results. We apply a magnetic field that is perpendicular to one of the four possible crystallographic axes. In this setup, we could eliminate the signal from
the NV centers with the other three crystallographic axes due to the detuning. So 
we perform a theoretical analysis only for the NV centers where the applied magnetic field is set to be orthogonal to the crystallographic axis.
The Hamiltonian of the NV center is described as follows
\begin{eqnarray}
H_{\rm NV }=  D{ \hat { S }  }_{ z }^{ 2 }
+ { E }_{ x } \left( { \hat { S }  }_{ x }^{ 2 }-{ \hat { S }  }_{ y }^{ 2 } \right)
+ { E }_{ y } \left( { \hat { S }  }_{ x }{ \hat { S }  }_{ y } + { \hat { S }  }_{ y }{ \hat { S }  }_{ x } \right)+g\mu _bB_x\hat{S}_x
\label{Hnv}
\end{eqnarray}
where $ \hat { S } $ is a spin-1 operator of the electron spin,
$ D $ is a zero-field splitting,
${ E }_{ x }({ E }_{ y })$ is a strain along $x(y)$ direction, and $g\mu _bB_x$ denotes a Zeeman splitting.
Also, without
loss of generality, we can set the magnetic field perpendicular to the crystallographic axis as the $x$ direction of the NV center.
The ground state is $|0\rangle $,
and we define the dark state and bright state as
$\ket{ D } = \frac { 1 } { \sqrt { 2 } } \left( \ket{ 1 } - \ket{ -1 } \right)$ and
$\ket{ B }  =\frac { 1 } { \sqrt { 2 } } \left( \ket{ 1 } + \ket{ -1 } \right)$, respectively. Under a condition of $D\gg g\mu _bB_x\gg E_y$, we can rewrite the Hamiltonian as 
\begin{eqnarray}
 H\simeq D'\hat{S}_z^2 +E_x'(\hat{S}_x^2-\hat{S}_y^2)\label{effectiveh}
\end{eqnarray}
where $D'=D+\frac{3}{2}\frac{(g\mu _bB_x)^2}{D+E_x}$ and $E_x'=E_x+\frac{1}{2}\frac{(g\mu _bB_x)^2}{D+E_x}$, and this has the same form as the Hamiltonian without applied magnetic fields.
It is known that two dips are observed around 2.87 GHz in the
CW-ODMR with zero magnetic field. Such dip structures occur when
the external driving induces transitions
from a ground state $|0\rangle $ to the other energy eigenstates such as $|B\rangle $
or $|D\rangle $\cite{Fang2013a,Rondin2014,Matsuzaki2016,Mittiga2018}.
Similar two dips should be observed in the ODMR with our setup due to the form of the Hamiltonian described in Eq. \ref{effectiveh}.

\begin{figure*}[t]
\begin{center}
\includegraphics[width=18cm]{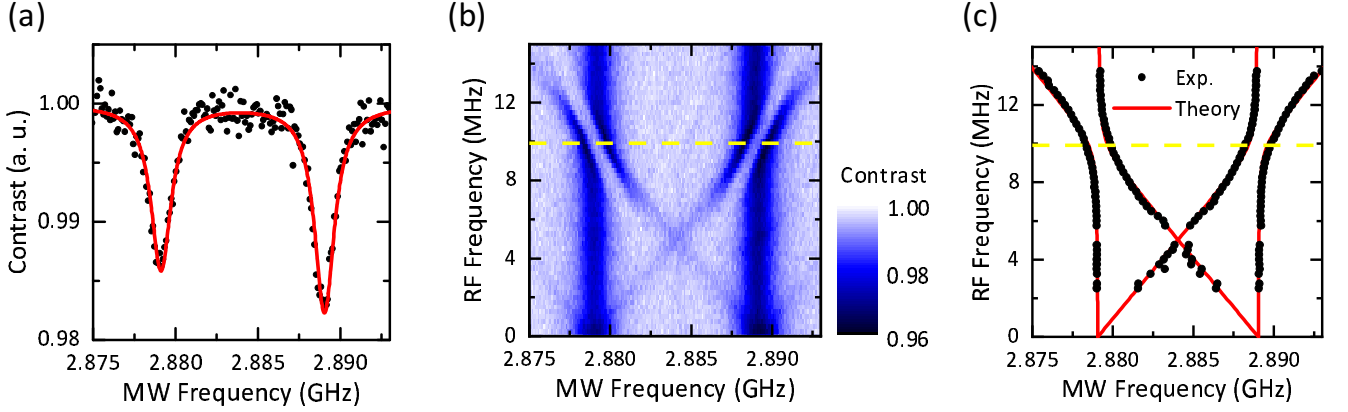}
\caption{(a) CW-ODMR spectrum with applied magnetic fields 
perpendicular to one of the four possible crystallographic axes. We observe resonances corresponding to $|B\rangle $ and $|D\rangle $ of the NV centers with an axis perpendicular to the magnetic fields, while other NV centers with different 
axes are 
well detuned by the magnetic fields.
(b) CW-ODMR with different frequencies of the microwave fields and  AC magnetic (RF) fields. The color contrast indicates the amount of photoluminescence. 
A clear anti-crossing structure is observed around $\ket{B}$-$\ket{D}$ resonance (shown by dotted line), and this is the manifestation of the RF-dressed states of $|B\rangle $ and $|D\rangle$.
(c) Comparison between the theory and experiment. We plot the resonant frequency observed in the experiment, and fit these data by our theoretical formula in the Eq. \ref{eq:allkai}.
There is a good agreement between them.
}
\abovecaptionskip=10pt
\label{f:ODMRwithBx}
\end{center}
\end{figure*}
Let us consider the dynamics of the NV centers with the driving fields. 
In our experiment, we performed the double resonance spectroscopy by simultaneously applying microwave and RF fields. Here, the RF fields corresponds to the target AC magnetic fields. The Hamiltonian of the driving fields is described by
\begin{eqnarray}
H_{\rm{ex}} = \sum_{j=x,y,z}{{\gamma}_eB}^{(j)}_{\rm mw} { \hat { S }  }_{ j }\cos { \left ({ \omega  }_{ \rm mw }t \right)  } + {{\gamma}_eB}^{(j)}_{\rm AC } { \hat { S }  }_{ j } \cos { \left({\omega}_{\rm AC}t \right) }
\label{Hex}
\end{eqnarray}
where ${\gamma}_e$ is the gyromagnetic ratio of the electron spin, ${ B }_{ \rm mw }({ B }_{ \rm AC })$ is a microwave (RF) fields amplitude, ${ \omega  }_{ \rm mw }({ \omega  }_{ \rm AC })$ 
is a microwave (AC magnetic) field driving frequency. 
The total Hamiltonian of the NV center with the driving fields is $ H = H_{\rm{nv}}+H_{\rm{ex}}$.
In a rotating frame with $U=-\omega_{\mathrm{AC}}\ket{D}\bra{D}-\omega_{\mathrm{mw}}\ket{0}\bra{0}$,  the effective Hamiltonian is
 \begin{align}
 H\simeq &(D-\omega_{\mathrm{mw}}+E_x)\ket{B}\bra{B}+(D-\omega_{\mathrm{mw}}-E_x+\omega_{\mathrm{AC}})\ket{D}\bra{D}+\nonumber\\
&+\frac{{{\gamma}_eB}^{(x)}_{\mathrm{mw}}}{2}(\ket{B}\bra{0}+\ket{0}\bra{B})+\frac{{{\gamma}_eB}^{(z)}_{\mathrm{AC}}}{2}( \ket{B}\bra{D} +\ket{D}\bra{B})
\end{align}
where we use  the rotating wave approximations.
Since we consider an ensemble of the NV centers, we can treat the system as a harmonic oscillator and so the Hamiltonian can be written as follows.
\begin{align}
H ^ { \prime } = \omega _ { b } \hat { b } ^ { \dagger } \hat { b } + \omega _ { d } \hat { d } ^ { \dagger } \hat { d } + J \left( \hat { b } ^ { \dagger } \hat { d } + \hat { b } \hat { d } ^ { \dagger } \right) 
+ \lambda _ { b } \left( \hat { b } + \hat { b } ^ { \dagger } \right)\label{eq:haizenH}
\end{align}
where $\omega _ { b } = D + E - \omega _ { \mathrm { mw } } , \omega _ { d } = D - E - \omega _ { \mathrm { mw } } + \omega _ { \mathrm { AC } } , J = \frac{{{\gamma}_eB}^{(z)}_{\mathrm{AC}}}{2} , 
\text { and } \lambda _ { b } = \frac{{{\gamma}_eB}^{(x)}_{\mathrm{mw}}}{2}$.
This Hamiltonian shows the states of \ket{B} or \ket{D} are coupled by
the applied RF
fields (that corresponds to the target AC magnetic fields), which results in RF-dressed states of the NV centers between \ket{B} or \ket{D}.
The Heisenberg equation, we obtain
\begin{align}
\begin{array} { l } { \frac { d \hat { b } } { d t } = - i \omega _ { b } \hat { b } - i J \hat { d } - i \lambda _ { b } - \Gamma _ { b } \hat { b } } \\ { \frac { d \hat { d } } { d t } 
= - i \omega _ { d } \hat { d } - i J \hat { b } - \Gamma _ { d } \hat { d } } \end{array}
\end{align}
where $\Gamma _ { b }$ $(\Gamma _ { d})$  denotes a decay rate of the state $|B\rangle $ ($|D\rangle $).
Since the system becomes steady in the CW-ODMR, we set $\frac { d \hat { b } } { d t } = \frac { d \hat { d } } { d t } = 0$ , and we obtain
 \begin{align}
 \begin{array} { l } { \hat { b } 
 = \frac { - \lambda _ { b } \left( \omega _ { d } - i \Gamma _ { d } \right) } { \left( \omega _ { b } - i \Gamma _ { b } \right) \left( \omega _ { d } - i \Gamma _ { d } \right) - J ^ { 2 } } } \\ 
 { \hat { d } = \frac { \lambda _ { b } J } { \left( \omega _ { b } - i \Gamma _ { b } \right) \left( \omega _ { d } - i \Gamma _ { d } \right) - J ^ { 2 } } } \end{array}
 \end{align}
The probability of the state to remain \ket{0} is described as $p_0=1-\hat { b } ^ { \dagger }\hat { b }-\hat { d } ^ { \dagger }\hat { d }$.
This probability corresponds to the amount of the photoluminescence in the ODMR.
From the expression, we can calculate the resonant frequency of the microwave as follows.
\begin{align}
\omega_\mathrm{mw}&=\frac{1}{2}\biggl\{2D\pm\omega_{\mathrm{rf}}\pm\sqrt{(2E_x-\omega_{\mathrm{rf}})^2+{{\gamma}_eB}^{(z)}_{\mathrm{AC}})^2} \biggr\}\label{eq:allkai}
\end{align}
and so there are four resonances in the ODMR which corresponds to AT splitting caused by RF field on the resonant excitation. We can calculate the sensitivity of the magnetic fields as follows.
\begin{eqnarray}
 \delta B_{\rm{AC}}^{(z)}=\frac{\delta S}{|\frac{dp_0}{dB_{\rm{AC}}^{(z)}}|}\label{sensitivitycal}
\end{eqnarray}
where $\delta S$ denotes the standard deviation of the signal.



Here, we describe our experimental setup to investigate the bandwidth of the AC magnetic field sensing with CW-ODMR.  A home-built system for confocal laser scanning microscopy was used, and the spatial resolution is around 400 nm.
We fixed a diamond sample above an antenna to irradiate the microwaves\cite{Sasaki2017}. 
 The target AC magnetic field was applied from a 30\,$\mu$m-diameter copper wire placed in contact with the sample surface. 
To apply DC magnetic fields perpendicular to one of the crystallographic axes of NV center, we used a magnet positioned under the antenna.
To measure the photons emitted from the NV centers, a single-photon resolving detector was used.

An ensemble of NV centers used in this experiment was in a 100\,nm-thick diamond film on a (001) electronic-grade substrate.
We had grown the isotopically  purified $^{ 12 }$C diamond film ([$^{ 12 }$C] =99.999\%) using nitrogen-doped microwave plasma-assisted chemical vapor deposition on the (001) electronic-grade substrate\cite{Balasubramanian2009,Watanabe2009}.
Since a high NV density and long coherence time is required for high sensitivity magnetic field sensing, we performed He$ ^{+} $ ion implantation at ion doses of $ { 10 }^{ 12 }$cm$^{ -3 }$ and 15 keV 
acceleration voltage, and then implemented annealing for 24 h in vacuum at $800^\circ$C to the sample\cite{Kleinsasser2016}. 
We measure the densities of NV and N, and they are of the order of $10^{15}$ and $10^{17}$ cm$^{-3}$\cite{Watanabe2009}.
About 40\% of the NV centers has a crystallographic axis with the same direction in our sample.
\begin{figure*}[ht]
\begin{center}
\includegraphics[width=18cm]{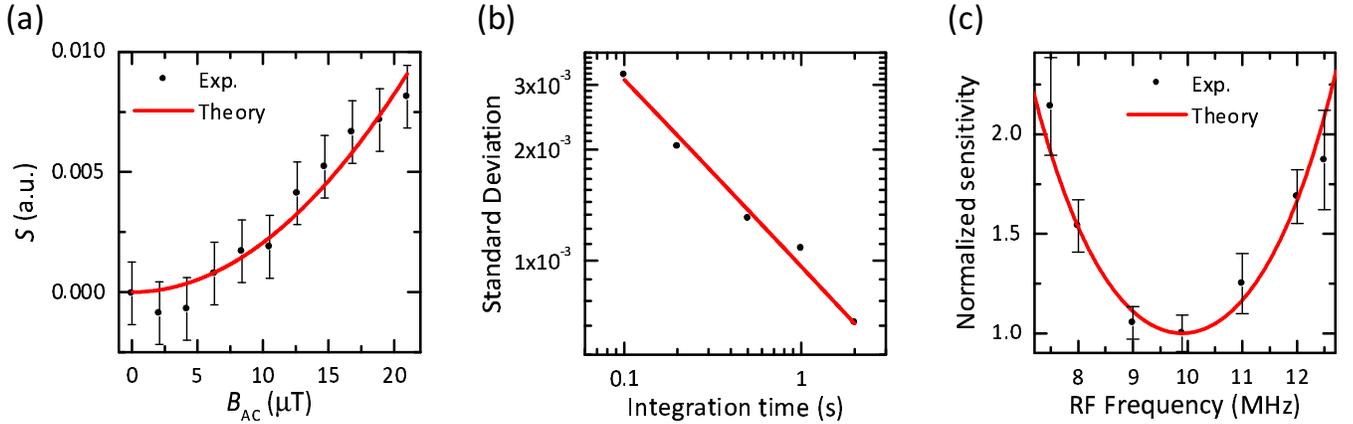}
\caption{(a) The optical signal strength plotted against the  amplitude of the AC magnetic fields (RF) in the CW-ODMR. Here, the frequency of the AC magnetic fields is fixed at $9.9$ GHz.
The signal shows quadratic dependence on the amplitude, and this is consistent with the results in \cite{saijo2018}.
(b) Plot of the standard deviation of the signal in the CW-ODMR against time. 
(c) Plot of the sensitivity of the AC magnetic fields sensing with CW-ODMR against the magnetic fields frequency. The bandwidth is estimated to be approximately 5 MHz
}
\abovecaptionskip=10pt
\label{f:RFbandwidth}
\end{center}
\end{figure*}

We explain our results.
Firstly, we measured the ODMR without AC magnetic field where we sweep only microwaves as shown in Fig. \ref{f:ODMRwithBx}(a). 
We observe two resonance in this frequency range as expected from the form of the Hamiltonian described in the Eq. \ref{effectiveh}, and these resonances indicate the transitions from $|0\rangle $ to $|B\rangle $ and $|D\rangle$. 
From this result, the frequency difference between $|B\rangle $ and $|D\rangle$ is determined as 9.9 MHz. 

Secondly, we performed the double resonance experiment.
More specifically, we measured the ODMR by applying both microwave and RF (AC magnetic) fields where we sweep both frequencies of microwaves and RF, as shown in the Fig. \ref{f:ODMRwithBx}(b). 
The microwave field makes the transition between $|0\rangle $ and RF-dressed states, and the reduction of the photoluminescence occurs due 
to such a resonance.  
Importantly, we observed anti-crossing when the frequency of RF
is around $9.9$ MHz in the Fig. \ref{f:ODMRwithBx}(b), and this corresponds to the
 transition frequency between \ket{B} and \ket{D} with AT splitting caused by RF field. 
From the ODMR, we determine the resonant frequency by a Lorentzian fitting. We compare these experimental results with the analytical solution in the Eq. 8,
and we reproduce the experimental results by our theoretical calculations, as shown in Fig. 2(c). There is good agreement between experiment and theory, except small resonances observed in experiments where the RF frequency is below 4 MHz and the 
microwave frequency is above (below) 2.890 (2.877)GHz. These small resonances might come from the two photon process to violate the rotating wave approximation, and the understanding of these is left as future work.

Finally, we measured the bandwidth of the AC magnetic field sensing with the double resonance CW-ODMR. Here, the microwave frequency for the ODMR 
is fixed at $\omega_{\mathrm{mw}}=D+0.5\omega_{\mathrm{AC}}$ to measure the AC magnetic field with a frequency of $\omega_{\mathrm{AC}}$.
The sensitivity can be estimated from the signal change in ODMR and the signal fluctuations \cite{saijo2018}. 
Especially, we plot the signal change and fluctuation of $\omega_{\mathrm{AC}}= 9.9~\mathrm{MHz}$ in the Fig. \ref{f:RFbandwidth}(a,b), and the sensitivity is around 4.9 $\mu \rm T/ \sqrt{\rm Hz} $ in our setup, which is comparable with the sensitivity
in the previously demonstrated AC magnetic field sensing with CW-ODMR \cite{saijo2018}. Similarly, we estimate the sensitivity with different frequency of the target AC magnetic fields, and we plot them in the Fig. \ref{f:RFbandwidth}(c) where we normalized the 
sensitivity by that with the frequency of $\omega_{\mathrm{AC}}= 9.9~\mathrm{MHz}$. 
Also, we fit these experimental results with our theoretical formula described in the Eq. \ref{sensitivitycal} where we use $\Gamma _b=\Gamma _d=\Gamma $ as a fitting parameter. 
We have a good agreement
between the theory and experiment where we obtain $\Gamma \simeq 2$ MHz, which is roughly consistent with the linewidth observed in the ODMR in the Fig. \ref{f:ODMRwithBx}(a).
These results show the bandwidth of our AC magnetic field sensor is around $5$ MHz. The bandwidth significantly depends on the linewidth.

In conclusion, we analyze the bandwidth of the AC magnetic field sensing using CW-ODMR based on electronic spin double resonance of NV centers in diamond. 
We perform the CW-ODMR by applying both microwave fields and RF fields with various frequencies.  As a result, we observe anti-crossing structure in the spectrum, which agrees with our theoretical prediction. This corresponds to AT splitting of RF-dressed state.
Based on these results, we investigate the sensitivity for the AC magnetic field sensing for several frequencies.
We find that the bandwidth is around 5 MHz at the center frequency of 9.9 MHz. At the center frequency, a sensitivity is estimated to be 4.9 $\mu \rm T/ \sqrt{\rm Hz} $ in our setup. 
Our results pave the way to realize a practical AC magnetic field sensor using a simple CW-ODMR setup. Furthermore, we expect these results to be the basis of the application of the phenomenon using the coupling between \ket{B} and \ket{D}.

We thank H. Toida and K. Kakuyanagi for helpful discussions. This work was supported y CREST (JPMJCR1774) and by  MEXT KAKENHI (Grant Nos. 15H05868, 15H05870, 15H03996, 26220602, and 26249108).

\end{document}